\documentclass[12pt]{article}
\usepackage{amsmath,amsfonts,amscd}
\usepackage{bm}
\usepackage{graphicx}
\def\eg{{\em e.g.\, }}
\def\ie{{\em i.e.\, }}
\def\etc{{\em etc.\, }}
\def\d{\partial}
\def\bra{\langle}
\def\ket{\rangle}
\def\hc{\dagger}
\def\bvec#1{{\bm #1}}
\def\vF{\bvec{F}}
\def\vp{\bvec{p}}
\def\vP{\bvec{P}}

\def\vr{\bvec{r}}
\def\vR{\bvec{R}}
\def\vx{\bvec{x}}
\def\vX{\bvec{X}}
\def\vgamma{\bvec{\gamma}}

\def\Z{\mathbb{Z}} 
 
\def\Zp{{\Z}_p}
\begin{document}
\title{Quantum kinetic equation before and after Big Bang}
\author{M.V.Altaisky\\ 
{\small Joint Institute for Nuclear Research, Dubna, 141980, Russia; and }\\
{\small Space Research Institute RAS, Profsoyuznaya 84/32, Moscow, 117997, 
Russia}\\
{\small e-mail: altaisky@mx.iki.rssi.ru}
}
\date{Apr 23, 2009}
\maketitle
\begin{abstract}
The energy dissipation in a gas of structured objects, \eg molecules, 
is considered in density matrix formalism. It is shown that the 
macroscopic irreversibility of the kinetic processes can be considered 
as a consequence of the microscopic operator ordering. Our approach is free 
of any special assumptions on the space-time 
geometry, except for the general causality assumptions, so it can be applied 
to a wide variety of processes, from the cosmological processes at Big Bang 
stage till the energy dissipation in molecular gases. 
\end{abstract}

\section{Introduction}
Kinetic equations describe the evolution of the distribution 
function of matter in phase  space at the presence of particle collisions.
They lay the foundation of many cosmological 
models, describing the formation of present barionic matter from 
initial plasma. The problem with kinetic approach is that at the Plank 
times, $\tau \sim 10^{-44}sec$, the era of quantum gravity, the 
Riemann space-time itself did not exist, so neither the 
Boltzmann kinetic equation of the form 
\begin{equation}
\frac{\d f(\vX,\vP,t)}{\d t} + \vF \frac{\d f(\vX,\vP,t) } {\d \vP} 
+\frac{\vP}{m}\frac{\d f(\vX,\vP,t) } {\d \vX} = I_{col}[f], 
\label{bke}
\end{equation}
nor its relativistic generalisation is valid for that stage. 

The situation with the quantum mechanical description of the Early Universe 
is a little better. In quantum mechanics the evolution of quantum 
system is described as the evolution of the density operator $\hat\rho$ 
obeying the von Neumann equation 
\begin{equation}
\imath\hbar\frac{\d}{\d t} \hat \rho = [\hat H, \hat \rho] 
\label{me},
\end{equation}
where $\hat H$ is the Hamiltonian of the system. 
For the evolution equation \eqref{me} it does not mean whether 
the spacetime is continuous, differentiable, Archimedian. What is 
significant is the evolution: the existence of time $t$ and the 
dependence of the state on $t$ just means the 
evolution; if there is no evolution there is no change in quantum state.

It is important for the kinetic approach that the evolution 
equation for the classical distribution function in phase space 
$f(\vX,\vP,t)$ should be obtained as the classical limit $\hbar\to0$ 
of a more fundamental quantum equation \eqref{me} at the assumption of the 
existence of classical trajectory $\vX=\vX(t)$ for each particle. 
This is called the {\em macroscopic limit}.

In a purely quantum case, \ie when $\hbar$ cannot be treated as a 
small parameter, it is impossible to assign any trajectory $\vX(t)$ to 
a given particle. This is the case for dense plasma, lasers, 
microelectronics \etc This is also the case for quantum gravity.

The present paper considers the effect of general causality assumptions 
on the evolution of the density operator and the kinetic equation, 
resulting from this evolution in macroscopic limit. The general causality 
assumptions, discussed in \cite{CC2005,AltaiskyPEPAN2005}, are required 
for the construction of quantum field theory on a category space, 
which is not a manifold, \ie when continuity and differentiability 
are not provided. This is the case for the quantum gravity era of the Early 
Universe, and hence before the light-cone causality of Minkovskian space have been set by Big Bang a more general causality assumptions should hold. 
In the limit of classical non relativistic gas the general theory drives 
us back to known results for the Boltzmann equation and the Klimontovich 
method for molecular gases, but ultimately leads to irreversibility 
if the internal degrees of freedom are involved.

The remainder of this paper is organized as follows. In {\em Section \ref{dens:sec}} we remind the links between quantum description of hierarchic systems and 
causality. In {\em Section \ref{kin:sec}} a semi-quantum kinetic equation 
for a gas of structured particles is constructed. {\em Section \ref{osc:sec}} 
presents a toy model of a system of hierarchic quantum oscillators and 
imposes the operator ordering on this system. In {\em Conclusion} we summarize 
the basic ideas of our approach.   

\section{Density operator for hierarchic structures \label{dens:sec}}
Strictly speaking the von Neumann equation \eqref{me} holds only 
for {\em closed} systems -- but the only system which is absolutely 
closed is the Universe as a whole. Therefore there should be some 
methods to treat the system of our interest as {\em approximately} closed, 
when its interaction with the rest of the Universe can be neglected, 
or treated perturbatively. In this case the state vector of the whole 
Universe can be casted in a form 
\begin{equation}
|\psi\ket = \sum_{i\alpha} c_{i\alpha} |\phi_i\ket |\theta_\alpha \ket,
\label{sv}
\end{equation}      
where $|\phi_i\ket$ are the states of the system, 
$|\theta_\alpha \ket$ are the states of its {\em environment}, \ie 
the rest part of the Universe. To get a tractable
model the states of the environment are assumed to be mutually orthogonal 
\begin{equation}
\bra \theta_\alpha|\theta_\beta \ket = \delta_{\alpha\beta}.
\label{onc}
\end{equation}
 In view of 
this assumption the matrix elements of the density operator of the whole 
Universe 
$$\hat \rho = |\psi\ket\bra\psi|$$ in the basis of the states of system 
$\phi$ are 
\begin{equation}
\rho_{i'i}^\phi= \sum_\alpha \overline{c_{i\alpha}}c_{i'\alpha}= {\rm Tr}_\theta
\bra i'|\hat \rho|i \ket. \label{dm}
\end{equation} 
This means the probabilities of different states of a quantum system
$i$ are obtained by averaging over all states of its environment.

The mean value of arbitrary physical observable $A$ for such system 
is equal to the trace of the product $\hat A \hat \rho$ with respect 
to the states of the system $\phi$:
\begin{equation}
\bra A \ket = {\rm Tr} \hat A \hat \rho = \sum_{ji} A_{ij}\rho^\phi_{ji}.
\label{mo}
\end{equation}

Since it is impossible to account for  all degrees of freedom of the 
environment, what is done in practice is different from the summation 
\eqref{dm} with orthonormality condition \eqref{onc}. 
Namely, considering a diluted gas of diatomic molecules in coordinate 
representation \cite{Klim1974,Petrus1974}, one uses the Wigner function
\begin{equation}
\rho(\vR,\vP,\vr,\vr',t) = \frac{1}{(2\pi)^3}\int 
\rho_1(\vR-\frac{\hbar}{2}\vgamma,\vR+\frac{\hbar}{2}\vgamma,
\vr_1',\vr_2',t)e^{-\imath\vgamma \vP} d\vgamma, 
\label{wf}
\end{equation}
where  $\vR$ is the center of mass coordinate of the molecule, $\vP$ is its 
total momentum; $\vr=\vr_1-\vr_2$ is the internal degree of freedom -- 
the relative displacement of atoms in the molecule; 
$\rho_1(\vr_1,\vr_2,\vr_1',\vr_2',t)$ is the density matrix of diatomic 
molecule in coordinate representation. 

In the language of state vectors the Wigner function formalism corresponds 
to the change of basis \eqref{sv} for the density operator \eqref{dm} to 
a hierarchic one
\begin{equation} \hat\rho = |\psi\ket \bra\psi|,\quad 
|\psi\ket = \sum_{i_\alpha i_{\alpha-1}} c_{i_\alpha i_{\alpha-1}} |\phi_{i_\alpha}^\alpha \ket 
|\phi_{i_{\alpha-1}}^{\alpha-1} \ket,
\label{wf1}
\end{equation} 
where $(\alpha-1)$ denotes the next to the studied system $(\phi^\alpha)$ 
hierarchy level. (Say, in the case of diatomic gas $|\phi^{\alpha-1}\ket$ denotes 
the state vector of molecule, with $|\phi^{\alpha}\ket$ being the state of 
atom in this molecule.) The density matrix for the system $(\phi^\alpha)$ 
can be casted in the form \cite{Alt03IJQI}: 
\begin{equation}
\hat \rho^\alpha = {\rm Tr}_{\alpha-1} |\psi\ket\bra\psi|,
\end{equation}
where $|\psi\ket$ is given by \eqref{wf1}.

The hierarchic representation \eqref{wf1} poses the problem of ordering the 
operators acting at different hierarchy levels. The operator ordering 
rule ``the coarse acts the first'' was suggested by the author 
\cite{AltaiskyPEPAN2005}. This corresponds to general cosmological 
idea of cascade process of the Universe origin from vacuum:
\begin{equation}
|U_0\ket = a_0^\hc |0\ket, |U_0U_{0i}\ket = a_i^\hc a_0^\hc |0\ket 
=\{ |U_0\ket, |U_{0i}\ket \}, \ldots.
\end{equation}
\begin{figure}[ht]
\centering \includegraphics{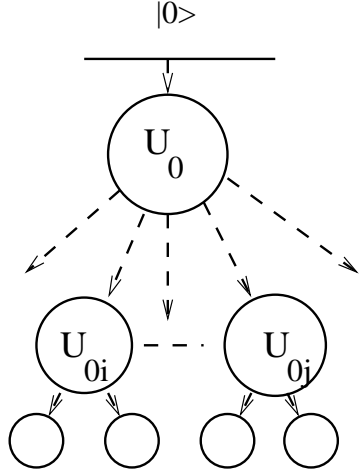}
\caption{Creation of the discrete Universe from vacuum as a cascade 
process}
\label{casc:pic}
\end{figure}
At present stage of the Universe evolution the hierarchy level and the number 
of constituents are very large $N \sim 2^p$ is expected to be of order of 
the Dirac-Eddington number $N\sim 10^{80}$, or so, and we observe continuous Universe \cite{BM1989,BGS2008}.

This implies two types of causality: ({\em i}) the vertical causality  $a\subseteq b$ ($a$ is a part of $b$); ({\em ii}) the horizontal causality 
$a\prec b$ ($a$ precedes $b$); in Fig.~\ref{casc:pic} 
$U_{0i} \subseteq U_0,\quad U_{0i}\prec U_{0j}$. Consequently, two types of 
{\em operator ordering} are required to meet two types of partial order. 
The approach with two types of causality relations is a generalization of 
common light-cone causality of the Minkovsky space with $T$-ordering 
relation for the operators on more general spaces. It perfectly 
meets the needs of quantum gravity level cosmological models, when the 
the manifold structure is not provided \cite{BLMS1987,BM1989}. The details of this topological 
approach, known as {\em region causality}, can be found in \cite{CC2005}.

What region causality says is quite natural not only for quantum gravity, 
but also for molecular physics: no experiment can detect a 
causal effect in a {\em point} -- what is really observed is an effect 
in a {\em region}, the size of which is restricted, at best, by the Heisenberg  
uncertainty relation $\Delta p \Delta x \ge \frac{\hbar}{2}$. So, in 
accordance to \cite{CC2005} we adopt two types of causality:
\begin{eqnarray*}
a\subseteq b & a\ \hbox{is a part of\ } b \\
a\prec b     & b\ \hbox{can see\ } a. 
\end{eqnarray*}   
The latter corresponds to the light-cone causality; the former, supplied 
with the operator ordering ``the coarse acts first'', just stands for the 
fact that it is impossible to change the state of the part without changing 
the state of the whole. 

It is important for kinetic theory that the Bogolubov causality 
relations for the scattering matrix \cite{Bog1955} 
\begin{equation}
\frac{\delta}{\delta g(x)} \left[
\frac{\delta S[g]}{\delta g(y)} S^\hc [g]
\right]=0, \quad x \stackrel{<}{\sim} y,
\label{bgc}
\end{equation}   
originally formulated for the light-cone causality 
-- $x<y$ means $x_0<y_0$ and $x \sim y$ means the interval between the 
events $x$ and $y$ is space-like interval -- without significant 
changes are generalized to the vertical causality as well. This can 
be easily seen by introducing $p$-adic coordinate $x\in \Zp$ on 
the branching tree. 
The research in quantum gravity have stimulated the extension of the 
$S$-matrix formalism to non-Archimedian numeric fields, namely to the 
field of $p$-adic numbers \cite{VVZ1994}. Hence the Bogolubov causality 
relation  \eqref{bgc} can be given for the quantum fields depending 
on $p$-adic arguments, and therefore the fields defined on the 
vertexes of $p$-adic tree can be ordered in analogy to the $T$-ordering 
causality.

In $p$-adic metric 
the partial order is given by $p$-adic metric $|\cdot|_p$: 

$x<y \hbox{\ if\ } |x|_p < |y|_p$ -- vertical ordering;

if  $|x|_p = |y|_p$ then $x$ and $y$ are ordered by the first nonzero 
coefficients -- this is horizontal ordering in the ring of natural 
numbers \cite{VVZ1994,Khrennikov1997}.   
So, if $x,y\in\Zp$ in \eqref{bgc} the Bogolubov causality relation 
is defined on the branching tree. A toy-model of such field theory 
is presented in \cite{AS1999}. 

In topology the causality of regions is axiomatising by means of partial order 
relations in a form of 
causal sets \cite{BLMS1987,Sorkin2003}.
The causality at quantum gravity level -- we would say before the Big Bang -- 
will be the {\em region causality} with two causality relations $A\subset B$  
and $A\prec B$ , that is called a causal site. The axiomatics of causal site given in \cite{CC2005} is presented in Appendix.

In a (pseudo-) Euclidean space the structure of the causal  sites implies causal 
paths and the geodesic coinciding with light-cone causality. However at 
quantum gravity level, when the Archimedean axiom does not hold, the 
vertical causality relation $\subseteq$ is expected to play the important  
role. 
We do not know the concrete scheme of the present Universe formation from 
initial object, but without loss of generality we can assume that at initial 
time $t=0$ there was only one initial object $U_0$. At the next instant of 
time $t$ it branched into $p\ge2$ parts, each of those continued further 
branching in a tree-like matter  -- therefore a hierarchic structure have been
formed \cite{AS1995}. This hierarchic structure endowed with certain system of relations 
between its elements has formed our space-time. 

\begin{figure}[h]
\centering \includegraphics[width=6cm]{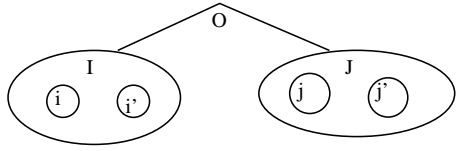}
\caption{Hierarchic binary system. Two means of calculating partial density 
matrices of the sub-parts can be applied}
\label{h3:pic}
\end{figure}

The axiomatics of causal site is rather general and should be applied not 
only at quantum gravity level, but also for quantum systems 
of present life, such as molecular gases. To see this, we have to 
cast the density operator in a hierarchic basis \eqref{wf1} .
Let us consider a hierarchic 
system $O$ consisting of two parts $I$ and $J$, each of those consists of 
its own two parts, $(i,i')$ and $(j,j')$, respectively, see Fig.~\ref{h3:pic}.
For definiteness, let us consider the particle ``i'' of the finest hierarchy 
level of the system above. The mean value of an observable $A$ 
related to $i$ can be evaluated in two different ways:
\begin{enumerate} 
\item with the density matrix averaged over all states of $i'$:
$$ \hat \rho_i = \sum_{i'} \bra i'| \hat \rho |i'\ket; $$
\item with the density matrix averaged over the states of $I$ -- the ``parent'' 
of $i$:
$$ \hat \rho_i = \sum_{I} \bra I| \hat\rho |I\ket. $$
\end{enumerate}
The former case is a standard way of the partial density matrix for a 
multi-particle system \cite{Klim1974}, the latter is presented in 
\cite{Alt03IJQI}.  In the 
latter case the state vectors are represented in a tree-like form 
\begin{equation}
|\Phi_I\ket =\left\{|I\ket ,   |Ii\ket, \ldots  \right\}, \quad 
\hat \rho = |\Phi_I\ket \bra \Phi_I|,
\end{equation}
with index $O$ being dropped as omnipresent.

To provide the equivalence of these two approaches the effect of the 
other particles $J\ne I$ on the parts of $I$, \ie on $i$,$i'$, should 
exist only via change the state of $I$. This the effect of the 
whole to its parts. In thermodynamics it corresponds to  
adiabatic insulation of the system $I$, when the energy can be transferred 
to the system as a whole, but not to its internal degrees of freedom.

\section{Quantum kinetic equation for the gas of oscillators \label{kin:sec}}
Let us consider a simple example. A homogeneous gas of molecules with the 
distribution function $f(\vP)$. The molecules can transfer part of their 
kinetic energy to their internal degrees of freedom, represented 
by quantum harmonic oscillators. The molecules do not interact beyond 
the interaction zone $|\vx_1-\vx_2|>2r_0$, with $r_0$ being the typical radius 
of the molecule. This approximation is valid since the energy of 
oscillatory degrees of freedom is much bigger than that of rotational 
degrees of freedom, but much less than the energy of electron transitions: 
\begin{equation}
E_{el} \gg E_{osc} \gg E_{rot}.
\label{eosc}
\end{equation}
So, in a certain range of energies, corresponding to infrared radiation 
energy, the kinetic energy of molecules is transferred into oscillatory degrees 
of freedom by means of binary collisions between molecules

The energy levels of quantum harmonic oscillator, representing 
the oscillatory degrees of freedom of the molecules, are 
\begin{equation}
E_n = \hbar \omega \left(n+ \frac{1}{2}\right), \quad n=0,1,\ldots
\label{osc}
\end{equation}
The state of each molecule in this semi-quantum approximation is 
given by three variables $(\vX,\vP,n)$ -- position, momentum and 
excitation number. The rotations are ignored in this approximation \eqref{eosc}. 
The gas is assumed to be homogeneous (to be in a constant force field), 
therefore the total momentum is conserved in each binary collision,  
the energy can be transferred to the internal degrees of freedom 
only by quanta $\Delta E = \hbar\omega$, but not in a continuous manner.
To make the model consistent we assume the oscillator degrees of 
freedom to be in thermodynamic equilibrium with radiation at temperature 
$T$. 

Taking into account that in quantum description \eqref{me} the density 
matrix can be casted in terms of the Wigner functions, 
we can derive the kinetic equation for the probability density functions 
$f_n(\vX,\vP,t)$ with the initial condition $f_{n>0}(\vX,\vP,0)=0$. 
The causality condition ``the coarse acts first'' will manifest itself 
in the restriction that the energy transfer from the macroscopic 
$(\vX,\vP)$ degrees of freedom  to the microscopic $n$ is allowed, 
while inverse is not: the excess of energy is relaxed into radiation only. 

Let us denote the momenta of two colliding molecules (in centre mass system) 
before the collision as $\vP$ and $\vP_1$, and  $\vP'$ and $\vP_1'$ 
after collision, respectively. 
The momentum and energy conservation laws for binary collision imply  
\begin{eqnarray}
\nonumber \vP+\vP_1&=&\vP'+\vP_1' \\
\frac{\vP^2}{2m}+\frac{\vP_1^2}{2m} + \hbar\omega(n+n_1) &=&
\frac{{\vP'}^2}{2m}+\frac{{\vP_1'}^2}{2m} + \hbar\omega(n'+n_1').
\end{eqnarray}
As a simplest model the assume the probability of transfer of $N$ quanta of 
energy   into internal 
degrees of freedom is proportional to the number of means $g(N)$ this 
energy can be distributed between two oscillators:
\begin{equation}
P(\Delta Q) = \frac{\sum_{N=0}^{N_{max}} g(N)\delta(\Delta Q -\Delta E_N)}{
\sum_{N=0}^{N_{max}} g(N)}, \quad 
g(N)=N+1,\quad \Delta E_N = \hbar\omega N.
\label{pht}
\end{equation}
The sums in \eqref{pht} should be bounded by some reasonable integer 
$N_{max}>0$, which guaranties the condition \eqref{eosc} holds. 

The collision integral for our model, \ie the change of particle density 
per unit of phase space volume, is given by 
\begin{eqnarray}
\nonumber I_{col}&=&\int d\Gamma_1d\Gamma' d\Gamma_1' w'(f'f_1'P(-\Delta Q)-ff_1P(\Delta Q)), \\
\nonumber d\Gamma &=& \frac{d\vX d\vP}{(2\pi\hbar)^3} ,\\
\nonumber 
w'&\equiv& w'(\vP',\vP_1'\to \vP,\vP_1) \sim \delta(\vP'+\vP_1'-\vP-\vP_1), \\
\nonumber f'&\equiv& f_{n'}(\vX',\vP',t), \\
\Delta Q &=& \frac{\vP^2}{2m}+\frac{\vP_1^2}{2m} -
\frac{{\vP'}^2}{2m}-\frac{{\vP_1'}^2}{2m}. \label{dq}
\end{eqnarray}
The detailed balance principle does not hold for the kinetic equation
\begin{equation}
\frac{\d f_n(\vX,\vP,t)}{\d t} + \vF\cdot \frac{\d f_n(\vX,\vP,t) } {\d \vP} 
+\frac{\vP}{m}\cdot\frac{\d f_n(\vX,\vP, t) } {\d \vX}
+\sum_l \kappa_{nl} f_l(\vX,\vP,t) 
= I_{col}[f] 
\label{ke1}
\end{equation}
for $P(\Delta Q) \ne P(-\Delta Q)$. The coefficients $\kappa_{nl}$ account for 
the interaction of the internal degrees of freedom with radiation. 

If the internal degrees of freedom, \ie the oscillators \eqref{osc}, are in 
thermodynamic equilibrium with radiation at temperature $T$, the ratios 
of the energy level populations are given by 
\begin{equation}
\frac{n_j}{n_k}=\exp\left(-\frac{E_j-E_k}{k_BT} \right),
\label{blz}
\end{equation}
where $n_j = \int f_j(\vX,\vP, t) d\Gamma$, $k_B$ -- is the Boltzmann constant;
for definiteness we assume $E_j > E_k$.    
In thermodynamic equilibrium the transition rate from $j$ to $k$ is equal to 
that from $k$ to $j$. Since the population of each energy level is kept stationary 
\begin{equation}
\frac{dn_j}{dt} = k_{jk}n_k - k_{kj}n_j=0,\quad k_{kj}=k_{jk},
\label{dnt}
\end{equation}
where the first term stands for the excitations $k\to j$ and the second -- for relaxations 
$j \to k$.

For molecular oscillators the typical frequencies are infrared 
$\omega_{jk} \sim 10^{14}\hbox{sec}^{-1}$. In dipole approximation (see \eg 
\cite{Flygare1978})
\begin{equation}
k_{jk}=\frac{4\omega_{jk}^3d_{jk}^2}{3\hbar c^3}\cdot 
\frac{1}{
\exp\left(\frac{\hbar\omega_{jk}}{k_BT} \right)-1
}, \label{kjk}
\end{equation} 
where $d_{jk}$ is the matrix element of dipole moment operator for the $j\to k$ 
transition. For molecular oscillators the equilibrium transition 
rate \eqref{kjk} has the order of magnitude $10^2 \hbox{sec}^{-1}$. 

For our case of the relaxation of the kinetic energy of macroscopic 
degrees of freedom into the microscopic oscillators the {\em spontaneous} 
and {\em stimulated} emission is of paramount importance. Spontaneous 
emission rate $A_{kj}$ is the probability of downward transition $j\to k$, 
extra to the equilibrium one $k_{kj}$ \eqref{dnt}.
In thermodynamic equilibrium the total numbers of upward and downward 
transitions are equal
\begin{equation}
k_{jk}n_k = (k_{kj}+A_{kj})n_j
\end{equation}
from where, with the symmetry relation $k_{jk}=k_{kj}$ and the equilibrium 
populations \eqref{blz}, the spontaneous emission rate is found to be
\begin{equation}
A_{jk}=\frac{4\omega_{jk}^3d_{jk}^2}{3\hbar c^3}. 
\label{Ajk}
\end{equation}
For the low temperatures, $\hbar\omega_{jk}\gg k_B T$, the spontaneous emission 
dominates over the equilibrium transitions $$A_{jk}\gg k_{jk}.$$ For typical 
molecular frequencies the order of magnitude of the spontaneous emission 
rate is as high as $A_{jk} \sim 40 \hbox{sec}^{-1}$ \cite{Flygare1978}.

If the time before collisions is much less than spontaneous emission time 
$$\tau_{col} \ll \frac{1}{A_{kj}},$$
which is often so in molecular collisions, than the macroscopic 
collision plays the role of the {\em measuring device} with respect to 
quantum oscillator degrees of freedom, \ie collisions force the spontaneous 
emission -- that is they result in stimulated emission. 

\begin{figure}
\centering \includegraphics{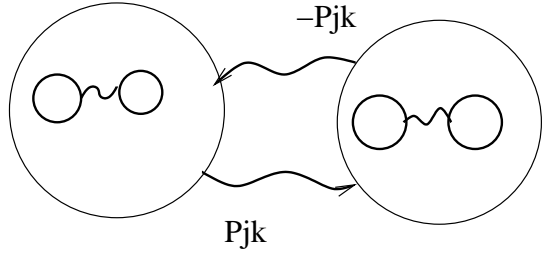}
\caption{Simultaneous energy relaxation by two molecules in a binary 
collision}
\label{rel2:pic}
\end{figure}

The {\em direct}, \ie the photon-less, energy transfer of the excitation 
energies of colliding molecules into their kinetic energy seems to be 
impossible. If the molecules are in excited state $j$ and during 
the collision they relax to the lower energy states $k$ the energy 
$E_j-E_k$ of the molecules is released in the form of photons 
with frequencies 
$$\omega_{jk}=\frac{E_j-E_k}{\hbar}$$
and the momentum  
$$\vp_{jk}=\frac{\omega_{jk}}{c}=\frac{E_j-E_k}{c},$$
see Fig.~\ref{rel2:pic}.
Due to the conservation of total momentum, the momentum of each molecule 
is changed by $-\vp_{jk}$, and its energy is increased by 
$$\Delta E \approx \frac{\vp_{jk}^2}{2m}$$ (in non-relativistic approximation, 
since momentum transfer is small).

Suppose now, the photons are virtual, rather than real. If so, 
the photon emitted by molecule 1 is absorbed by the molecule 2, 
and drives it back to the excited state j; the photon emitted by the 
molecule 2 is absorbed by the molecule 1 and drives it back to the 
excited state j. In such a way we have increased the kinetic energy 
of colliding molecules without changing their internal states. This 
is a kind of {\em perpetum mobile} and is therefore forbidden. In 
the language of hierarchic state vectors \eqref{wf1} such process of energy 
transfer from $|\phi^\alpha\ket$ to $|\phi^{\alpha-1}\ket$ is 
prohibited by the operator ordering rule ``the coarse $(\alpha-1)$ acts first''.  

Similarly, the molecules in Fig.~\ref{rel2:pic} may be two mesons in excited states. In 
that case the quark confinement prohibits the transfer of internal excitation 
energy into the meson kinetic energy.

\section{Oscillator toy model \label{osc:sec}}
In previous section we considered a gas of molecules where the kinetic 
energy of colliding particles is redistributed into internal excitation 
energy of particles by means of inelastic collisions. The particle 
motion in phase space was described classically, but the internal 
degrees of freedom were treated as quantum oscillators. It is possible 
to construct a toy hierarchic model treating where all 
degrees of freedom are quantized. 

\begin{figure}[ht]
\centering \includegraphics[height=2in]{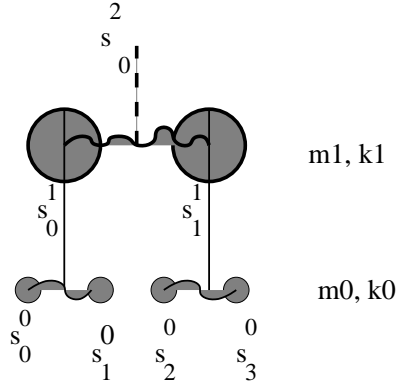}
\caption{Toy model system of hierarchic oscillators}
\label{hosc:pic}
\end{figure}

Let us consider a  system of equally spaced particles of mass $m_0$ located 
at $x=s^0_0,s^0_1,s^0_2,\ldots$, such that each $(2i)$-th particle is 
connected to its neighbour $(2i+1)$-particle by a harmonic potential of 
rigidity $k_0$. The hierarchic structure is 
constructed in a way that two ``atomic'' particles. $(2i, 2i+1)$, 
form the $i$-th molecule of mass $m_1=2m_0$; the molecules, in their 
turn interact to each other in the same hierarchic way as atoms, 
but with different rigidity constant $k_1$, see Fig.~\ref{hosc:pic}.   
The Hamiltonian of the hierarchic two-level system shown in Fig.~\ref{hosc:pic} 
has the form 
\begin{equation}
H= \frac{m_0}{2} \sum_{i=0}^3 \left(\dot {s^0_i} \right)^2 + 
\frac{k_0}{2} \sum_{i=0}^1 \left(d^1_i\right)^2
+\frac{k_1}{2}\sum_{i=0}^0 \left(d^2_i\right)^2,
\end{equation} 
where we have introduced the hierarchic centre mass coordinates 
\begin{equation}
s^{j+1}_i = \frac{s^j_{2i}+s^j_{2i+1}}{2} \label{hcm}
\end{equation}
and the displacements 
\begin{equation}
d^{j+1}_i = s^j_{2i}-s^j_{2i+1} \label{hds};
\end{equation}
so that the inverse transform is 
\begin{equation}
s^j_{2i} = s^{j+1}_i+ \frac{d^{j+1}_i}{2},\quad 
s^j_{2i+1} = s^{j+1}_i- \frac{d^{j+1}_i}{2}.
\end{equation}
The straightforward algebra gives 
$$
H = \frac{m_0}{2} \sum_{i=0}^1 2 \left(\dot s^1_i \right)^2
  + \frac{m_0}{2} \sum_{i=0}^1 \frac{1}{2}  \left(\dot d^1_i \right)^2
  + \frac{k_0}{2} \sum_{i=0}^1 \left(d^1_i \right)^2
  + \frac{k_1}{2} \sum_{i=0}^0 \left(d^2_i \right)^2,
$$
or, after simplification,  
\begin{eqnarray*}
H = m_0 \left[
\left(\dot s^2_0+\frac{\dot d^2_0}{2}\right)^2 +
\left(\dot s^2_0-\frac{\dot d^2_0}{2}\right)^2
\right] + \frac{k_1}{2} \left(d^2_0 \right)^2
+ \frac{m_0}{4} \sum_{i=0}^1 \left(\dot d^1_i \right)^2
+ \frac{k_0}{2} \sum_{i=0}^1 \left(d^1_i \right)^2.
\end{eqnarray*}
Since $\dot s^2_0=0$ because of the zero total momentum, the final 
equation for the Hamiltonian is a sum of two independent oscillators, 
belonging to ``0'' and ``1'' scales, respectively
\begin{equation}
H = m_0 \frac{\left(\dot d^2_0\right)^2}{2} + \frac{k_1}{2} \left( d^2_0\right)^2 +  \frac{m_0}{4} \sum_{i=0}^1 \left(\dot d^1_i \right)^2 + \frac{k_0}{2} \sum_{i=0}^1 \left(d^1_i \right)^2 .
\label{fh} 
\end{equation}
For each level of the hierarchy it is possible to substitute creation 
and annihilation operators $a^\hc,a$ and the number of excitation. In this 
way the scattering of two molecules described by $d^2_0$ coordinate 
will result in relaxation of the excitation and redistribution of this energy 
into the modes $d^1_0,d^1_1$, so that 
\begin{equation}
\hbar \omega_2 \Delta n^2_0 = \hbar \omega_1 (\Delta n^1_0 + \Delta n^1_1).
\label{ecl}
\end{equation}

In classical toy models, like the Toda chains of oscillators there 
is no such independence of oscillations of different scales as in our 
model considered above. If the particles are allowed to interact with 
{\em all} the neighbours, including that in the next block, the interaction Hamiltonian 
will have an extra term. Say, for the system shown in Fig.~\ref{hosc:pic} 
this will be the term $$H_{ib}=\frac{\tilde k_0}{2}\sum_i (s^0_{2i+1}-s^0_{2i+2})^2.$$
For the first two blocks this gives the contribution 
\begin{equation}
\Delta H_{ib} = \frac{\tilde k_0}{2}(s^0_{1}-s^0_{2})^2=\frac{\tilde k_0}{2}\left(d^2_0-\frac{d^1_0+d^1_1}{2} \right)^2, \label{hib}
\end{equation}
which produces the inter-scale energy transfer. 

The Hamiltonian \eqref{fh} does not contain any cross-scale interaction 
terms. Therefore the equation \eqref{ecl} is just a formal equation 
for energy conservation without any specification of the processes 
of energy exchange. If the cross-scale interaction \eqref{hib} is 
introduced, the free oscillators can be casted in terms of creation 
and annihilation operators $a^\hc,a$ and $A^\hc,A$
\begin{equation}
d_0^2 = \sqrt{\frac{\hbar}{\sqrt{m_0k_0}}} \frac{A+A^\hc}{2}, \quad 
d_i^1 = \sqrt{\frac{\hbar}{\sqrt{\mu_0 k_0}}} \frac{a_i+a^\hc_i}{2}, \quad i=0,1,
\end{equation}
where $\mu_0 = m_0/2$. The operators $a$ and $a^\hc$ are related to the 
normal modes $\Omega=\sqrt{k_1/m_0}, \omega = \sqrt{k_0/\mu_0}$ 
coordinate and momenta  
$$
d^1_i = \sqrt{\frac{\hbar}{\mu_0 \omega}}\frac{a_i+a^\hc_i}{2},\quad
p^1_i= \sqrt{\hbar\mu_0\omega}\frac{a_i-a^\hc_i}{2\imath}.$$ 
The inter-scale interaction Hamiltonian \eqref{hib} in second quantization 
formalism has the form
\begin{equation}
H_{ib} = \frac{\tilde k_0 \hbar}{2}\Bigl[
\frac{1}{(k_1m_0)^{1/4}}\frac{A+A^\hc}{2} 
-\frac{1}{2}\frac{1}{(k_0\mu_0)^{1/4}}\sum_{i=0}^1\frac{a_i+a^\hc_i}{2} 
\Bigr]^2
\end{equation}
If we assume the scale ordering -- an analog of $T$-ordering, -- 
then only the terms of the form $a_i^\hc A$ will contribute 
to cross-scale interaction. 

\section{Conclusion}
\begin{figure}[h]
\centering \includegraphics[width=12cm]{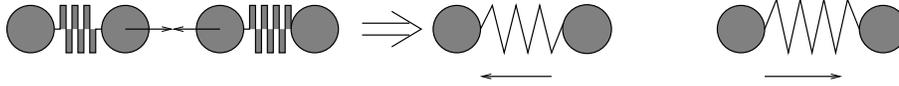}
\caption{Classical accumulation of mechanic energy: the potential energy 
of contracted springs of two oscillators can be relaxed in the form of kinetic 
energy of scattered oscillators}
\label{sc:pic}
\end{figure}

In classical theory of molecular gases the kinetic energy of molecules 
can be transformed into the heat energy of their internal degrees of freedom 
and vice versa: the energy accumulated in the internal degrees of freedom can 
be relaxed into kinetic energy of scattering particles, see Fig.\ref{sc:pic}.
In purely quantum case, when momentum and position become operators, 
the direct energy transfer from internal (microscopic) to macroscopic 
degrees of freedom seems to be impossible. 

The mathematical reason for that is the need to order the operators 
related the part and the whole (the micro and the macro) degrees of freedom.
Apart from the requirement of finiteness, which is not discussed in this paper,
this need stems from the fact that the state of the part cannot be 
affected unless the state of the whole is affected (measured) first. 
This is axiomatized by the operator ordering rule ``the coarse 
acts first'' \cite{AltaiskyPEPAN2005}. The set of axioms with two 
partial ordering operations \cite{CC2005}, aimed for the development of 
quantum gravity theory, may or may not be self-evident at common 
scales $l\gg l_{Pl}$. However, there exists a simple example  illustrating 
the impossibility  of the process, shown in Fig.~\ref{sc:pic}, for a pair 
of quantum oscillators.

Indeed, in classical system the potential energy is accumulated into the 
springs by decreasing the size of oscillators, the relaxation of potential 
energy into kinetic energy of scattered particles goes through the expansion 
-- the increase of oscillator sizes. For the quantum oscillator in the 
$n$-th excited state the mean squared size is 
$$
\bra \xi^2 \ket_n = n+\frac{1}{2}, \quad E_n = \hbar\omega \bra \xi^2 \ket_n.$$
Thus the relaxation from high to low energy level results in ``contraction'' 
of oscillator, rather than in expansion. So, no scattering takes place unless 
an extra momentum is injected. 

In the final end the axiomatizing of operator ordering used above provides 
a reformulation of the second thermodynamics's law on the microscopic level. 
For a microscopic system, described classically, the heating of internal 
degrees of freedom cannot result in macroscopic movement if it is accompanied 
by the decrease of entropy. In quantum case the decrease of entropy can be 
prevented by operator ordering. 

\section*{Acknowledgement}
The author is thankful to Prof. V.B.Priezzhev and Dr. B.G.Sidharth 
for useful comments. The paper was supported  by DFG Project 436 RUS 113/951.
\appendix
\section{Axiomatic structure of causal sites}
A {\em causal site } is a set of ``regions'' with two binary relations 
denoted $\subseteq$ and $\prec$ satisfying the axioms:\\
1. $\forall A,B,C$\\
(a) $A\subseteq B \wedge B \subseteq C \Rightarrow A\subseteq C$,\\ 
(b) $A\subseteq A$, \\
(c) $A\subseteq B \wedge B \subseteq A \Rightarrow A=B$.\\
2. The partial order $\subseteq$ has a minimum element $\phi$, called 
{\em an empty region}.\\
3.The partial order  $\subseteq$ has {\em unions}: $\forall A,B \exists A\cup B$, so that\\
(a) $A\subseteq A\cup B \wedge B \subseteq A\cup B$ \\
(b) $A\subseteq C \wedge B \subseteq C \Rightarrow A\cup B \subseteq C$.\\
4. The partial order  $\prec$ induces a strict partial order on the nonempty  
regions:\\
(a) $A\prec B \wedge B \prec C \Rightarrow A\prec C$,\\
(b) $A\not\prec A$ \\
5. $\forall A,B,C$\\
(a)$A\subseteq B \wedge B\prec C \Rightarrow A \prec C$,\\
(b)$A\subseteq B \wedge C\prec B \Rightarrow C \prec A$,\\
(c)$A\prec C \wedge B\prec C \Rightarrow A\cup B \prec C$.\\
6. $\forall A,B \exists B_A$, cutting of $A$ by $B$, so that \\
(a) $B_A \prec A \wedge B_A \subseteq B$,\\
(b) $D\prec A \wedge D \subseteq B \Rightarrow D \subseteq B_A$.\\
7. If $A\prec C$ are nonempty regions, and $\exists D: A\prec D \prec C$, 
then $\exists B$ complete with respect to $A\prec C$ 

Def.: If $A\prec B \prec C$, then $B$ is said to be {\em complete} with respect 
to a causal pair $A\prec C$,  
if any causal path from $A$ to $C$ can be refined to a causal path from $A$ to $C$, on of whose members is contained in $B$.

Def.: A {\em causal path } $P$ is a sequence of nonempty regions $A_1 \prec A_2 \prec \ldots \prec A_n$. 


\end{document}